\documentstyle[12pt,epsfig]{article}

\begin{document}

\def\b{\bar}
\def\d{\partial}
\def\D{\Delta}
\def\cA{{\cal A}}
\def\cD{{\cal D}}
\def\cK{{\cal K}}
\def\f{\varphi}
\def\g{\gamma}
\def\G{\Gamma}
\def\l{\lambda}
\def\L{\Lambda}
\def\M{{\cal M}}
\def\m{\mu}
\def\n{\nu}
\def\p{\psi}
\def\q{\b q}
\def\r{\rho}
\def\t{\tau}
\def\x{\phi}
\def\X{\~\xi}
\def\~{\tilde}
\def\h{\eta}
\def\bZ{\bar Z}
\def\cY{\bar Y}
\def\bY3{\bar Y_{,3}}
\def\Y3{Y_{,3}}
\def\z{\zeta}
\def\Z{{\b\zeta}}
\def\Y{{\bar Y}}
\def\cZ{{\bar Z}}
\def\`{\dot}
\def\be{\begin{equation}}
\def\ee{\end{equation}}
\def\bea{\begin{eqnarray}}
\def\eea{\end{eqnarray}}
\def\half{\frac{1}{2}}
\def\fn{\footnote}
\def\bh{black hole \ }
\def\cL{{\cal L}}
\def\cH{{\cal H}}
\def\cF{{\cal F}}
\def\cP{{\cal P}}
\def\cM{{\cal M}}
\def\ol{\overline}
\def\const{{\rm const.\ }}
\def\ik{ik}
\def\mn{{\mu\nu}}
\def\a{\alpha}

\title{Spinning Particle, Rotating Black Hole and
Twistor-String
\footnote{Talk at the SPIN 2004 Symposium at Trieste. }}

\author{Alexander Burinskii\\
\\
 NSI Russian Academy of Sciences\\
B. Tulskaya 52, 115191 Moscow, Russia}
\maketitle

\begin{abstract}Structure of the spinning particle based on the
rotating black hole solution is considered.
It has gyromagnetic ratio $g=2$ and a nontrivial twistorial and
stringy systems.
The mass and spin appear from excitations of the Kerr circular string,
while the Dirac equation describes excitations of an {\it axial} stringy
system which is responsible for scattering.
Complex Kerr geometry contains an open twistor-string, target space of
which is equivalent to the Witten's `diagonal'
of the $\bf CP^3\times CP^{*3}$.
\end{abstract}

\section*{Rotating Black Hole as Spinning Particle}
In this paper we consider the model of spinning particle \cite{BurTwi}
which is based on the Kerr rotating black hole solution and has a reach
spinor, twistor and stringy structures.
The inspired by strings and twistors methods have led to
the strong progress in computation
of some scattering amplitudes  \cite{BDK}.
In the recent paper \cite{WitTwi} Witten suggested a
`twistor-string' which may be an element of the structure of  fundamental
particles. We show
that a version of the `twistor-string' is presented in the Kerr spinning
particle.

{\bf Twistors}  may be considered as  the lightlike world-lines
\cite{Pen}. The lightlike momentum $p^\m$ may be represented
in the spinor form $ p^\m = \bar \psi \sigma ^\m \psi \ .$ Twistor
is a generalization of the spinor which is necessary for description of
the null word-lines  carrying an angular
momentum, $x^\m (t)= x_0^\m + p^\m t$ with $x_0\ne0$. It contains
two extra spinor components
$\omega _{\dot \alpha}= \psi ^\alpha x_0^\n \sigma _{\n \alpha \dot
\alpha}$. The set $\{ x^\m, \ \psi ^\alpha \} $,  or the equivalent
set $\{ x^\m, \ Y \} $, where $Y = \psi _2 / \psi _1 $  is a
projective spinor, may also be considered as twistors.

\begin{figure}[ht]
\centerline{\epsfxsize=2.1in\epsfbox{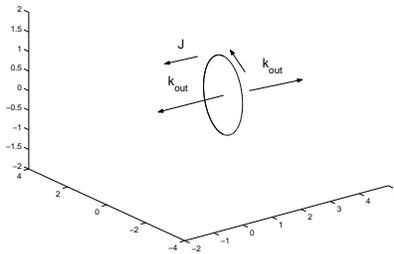}}
\caption{Kerr's singular ring and two semistrings (twistors)
of opposite chiralities. \label{kout}}
\end{figure}

{\bf Black Hole which is neither `Black' nor `Hole'} - this joke by
P. Townsend has direct relation to the Kerr spinning particle.
The ratio angular momentum/mass, $a=J/m$, for spinning particles is
very high and the black-hole horizons disappear revealing the naked
singular ring which is the branch line of the Kerr space on the
physical and `mirror' sheets.
The strings and twistors going through the Kerr ring pass into a `mirror'
world and look as semi-infinite.
 In \cite{Bur0} a model of spinning particle was
suggested, where the quantum electromagnetic excitations of the Kerr
ring generate the spin and mass. It was recognized that the
Kerr ring is a closed string with traveling waves \cite{BurOri}, and
excitations  of the Kerr ring lead to the appearance of the
extra {\it axial} stringy system consisting of two semi-infinite
singular strings of opposite chiralities \cite{BurTwi}, see
Fig.~\ref{kout}.
These excitations generate the chiral
traveling waves along the axial semistrings, and the Dirac spinor
$\Psi = \left(\begin{array}{c}
\phi _\alpha \\
\chi ^{\dot \alpha}
\end{array} \right)$
  describes in the Weyl basis an interplay of
two axial traveling waves of opposite chiralities \cite{BurTwi}.
This axial stringy system plays important role in the formation of
the third stringy structure, the {\it complex} twistor-string.
\bigskip

{\bf Basic properties of the Kerr spinning particle:}
\begin{itemize}
\item Anomalous gyromagnetic ratio $g=2$ as that of the Dirac electron,
\item Twistorial and stringy structures,
\item Mass and spin of the particle appear from electromagnetic excitations
of the Kerr ring (traveling waves) - the Wheeler's `geon',
\item Compton region is structured by the Kerr circular string.
\item Dirac equation describes traveling waves on two axial semistrings.
\end{itemize}
A few properties may also have relation to foundations of quantum theory:

- Wave function has a physical carrier - the axial stringy system.

- Axial string may control the motion of particle due to a
topological coupling to circular string, which reproduce the old de
Broglie conjecture.

- The quantum property - absence of radiation by oscillations - is exhibited
here at the classical level due to the twofoldedness of
the Kerr space: the loss of energy on the `physical' sheet of space
 is compensated by ingoing radiation from the `mirror' sheet.

Besides, there are the relations to the Skirme and chiral bag models:
the Kerr congruence is a twisting generalization of the `hedgehog' ansatz.
\section*{Complex Kerr geometry and twistor-string}
In the {\bf Kerr-Schild formalism} \cite{DKS}
metric has the form $ g_{\m\n}
= \h_{\m\n} + 2 h k_{\m} k_{\n},$
where $ \h_\mn $ is the metric of Minkowski space, $x^\m =(t,x,y,z),$
 and $k_\m$ is a vortex of the
null field which is tangent to the Kerr's congruence of twistors
$\{x^\m, \ Y \}$ which is determined by the {\bf Kerr
theorem}\cite{Pen,BurTwi}. Congruence of twistors, a geodesic family
of null rays, covers the Kerr space twice (see figures in
\cite{BurTwi,BurOri}.) Two axial semistrings $z^-$ and $z^+$ are created
by two twistors corresponding to $Y=0$ and $Y=\infty$.

The complex Kerr string appears naturally in the Newman-initiated
{\it complex representation} of the Kerr geometry \cite{New} which
is generated by a complex world-line $X_0^\m (\t) \in {\bf CM^4}.$
The complex time $\t=t_0+i\sigma$  forms a stringy world-sheet
\cite{OogVaf,BurStr}. The real fields on the real space-time $x^\m$
are determined via a complex retarded-time construction, where the
vectors $K^\m=x^\m - X_0^\m(\t)$ have to satisfy the complex
light-cone constraints $K_\m K^\m =0$. This allows one to select two
families of twistors: left (holomorphic) $\{ X_0, \ Y \}$ and right
(antiholomorphic) $\{\bar X_0, \ \bar Y \}.$

In the Kerr case $X_0^\m(\t)=(\t,0,0,ia),$ and the complex
retarded-time equation $ t-\t \equiv t-t_0 -i\sigma =\tilde r $
(where $\tilde r = r+i a \cos \theta$ is the Kerr complex radial
distance and $\theta$  is an angular direction of twistor) shows
that on the
real space $\sigma= - a \cos \theta .$  Therefore, the light-cone
constrains select a strip on $\t$ plane, $\sigma \in [-a,+a]$, and
the complex world-sheet acquires the boundary, forming {\it an open
string} $ X_0^\m(t,\sigma).$
\begin{figure}[ht]
\centerline{\epsfxsize=3.7in\epsfbox{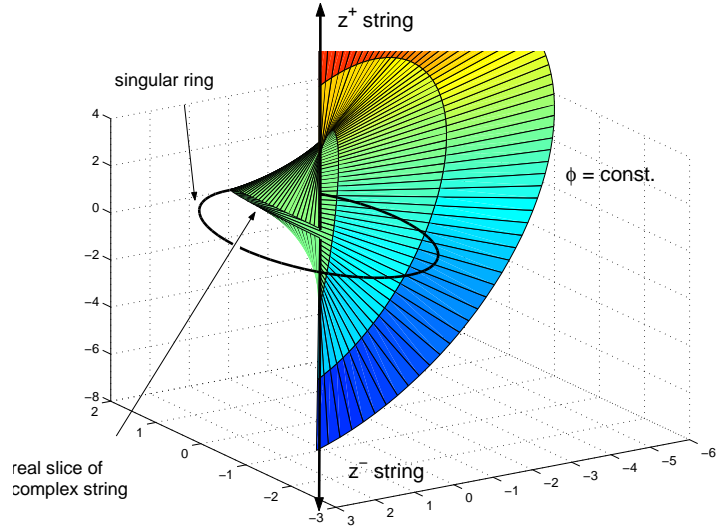}} \caption{The
complex twistor-string is stuck to two semistrings by imbedding into
the real Kerr geometry. The subset of semi-twistors at $\phi=const.$ is
shown. \label{twstr}}
\end{figure}

This string is similar to the well known $N=2$ string \cite{OogVaf}.
In ${\bf CM^4}$ it has only the chiral left modes $X_0(\t)$, while
the complex conjugate string has only right ones, $\bar X_0(\bar
\t).$ By imbedding this string into the real space-time, the left
and right structures are identified by orientifolding the
world-sheet \cite{BurStr}. The resulting string describes a massive particle and
has a broken N=2 supersymmetry. Its target space is equivalent to
`diagonal' of $\bf CP^3\times CP^{*3}$.

The twistors, joined to the ends of the complex string $X_0^\n
(t\pm ia)$, have the directions $\theta= 0, \ \pi$ and are generators
of the singular $z^\pm$ semi-strings, so {\it the complex string
turns out to be a D-string which is stuck to two singular
semistrings of opposite chiralities}, see Fig.~\ref{twstr}.
These $z^\pm$ singular semistrings may carry the Chan-Paton factors
(currents) playing the role of quarks with respect to the
complex string.
The Kerr circular string is responsible for the mass and spin of the
particle, while the axial semistrings are responsible for
the scattering processes.
\subsection*{Acknowledgments}
Author thanks the ICTP and Organizers of the SPIN2004 Symposium for
invitation and financial support, and the RFBR for travel
grant. Work is partially
supported by grant 03-02-27190 of the RFBR.


\begin{thebibliography}{0}
\bibitem{BurTwi} A. Burinskii, {\it Phys. Rev.} D{\bf 70} 086006 (2004),
hep-th/0406063.

\bibitem{BDK} Z. Bern, L. Dixon and D. Kosower,
{\it Ann. Rev. Nucl. Part. Sci.}{\bf 46}, 109 (1996);
V. P. Nair, {\it Phys. Lett.}{\bf B214}, 215 (1988).

\bibitem{WitTwi}
E. Witten, hep-th/0312171.
\bibitem{Pen} R. Penrose, {\it J. Math. Phys.}{\bf 8}, 345 (1967).
\bibitem{Bur0}  A. Burinskii, {\it Sov. Phys. JETP} {\bf 39} 193 (1974).

\bibitem{BurOri} A. Burinskii,
{\it Phys. Rev.} D{\bf 68}, 105004 (2003).

\bibitem{DKS}  G. C. Debney, R. P. Kerr, A. Schild, {\it J. Math.
Phys.}{\bf 10}, 1842 (1969).
\bibitem{New} E.T. Newman, {\it J. Math. Phys.} {\bf 15}, 44 (1974);
{\it Phys. Rev.} D{\bf 65}, 104005 (2002).

\bibitem{OogVaf} H. Ooguri, C. Vafa, {\it Nucl.
Phys.}{\bf B361}, 469 (1991); {\bf B367}, 83 (1991).
\bibitem{BurStr}
A. Ya. Burinskii, {\it Phys. Lett.}{\bf A185}, 441 (1994); gr-qc/9303003.

\end{thebibliography}
\end{document}